\documentclass[10pt,twocolumn]{article}

\usepackage{graphicx}
\usepackage{amsfonts}
\usepackage{amsmath}
\usepackage{amssymb}
\usepackage{mathrsfs} 
\usepackage{lipsum}
\usepackage{color}
\usepackage{siunitx}
\usepackage{authblk}
\usepackage[margin=2cm]{geometry}
\usepackage{mathtools, cuted}

\usepackage{bm}
\usepackage{amsmath}

\title{Natural convection above circular disks of evaporating liquids}
\author[1]{Benjamin Dollet}
\author[2]{Fran\c{c}ois Boulogne}
\affil[1]{Institut de Physique de Rennes, UMR 6251 CNRS and Universit\'e Rennes 1, Campus Beaulieu, B\^atiment 11A, 35042 Rennes Cedex, France}
\affil[2]{Laboratoire de Physique des Solides, CNRS, Univ. Paris-Sud, Universit\'e Paris-Saclay, Orsay 91405, France}

\date{\today}
\begin{document}

\twocolumn[
    \begin{@twocolumnfalse}
        \maketitle
        \begin{abstract}
            We investigate theoretically and experimentally the evaporation of liquid disks in the presence of natural convection due to a density difference between the vapor and the surrounding gas.
            From the analogy between thermal convection above a heated disk and our system, we derive scaling laws to describe the evaporation rate.
            The local evaporation rate depends on the presence of a boundary layer in the gas phase such that the total evaporation rate is given by a combination of different scaling contributions, which reflect the structure of the boundary layer.
            We compare our theoretical predictions to experiments performed with water in an environment controlled in humidity, which validate our approach.
        \end{abstract}
    \end{@twocolumnfalse}
]

%
%
\section{Introduction}
The evaporation of small liquid disks takes its roots in botany, in particular with some studies on the transpiration of plants published in the early twentieth century \cite{Livingston1912,Thomas1917,Thomas1917a,Jeffreys1918}.
Beyond this original inspiration, evaporation is an ubiquitous phenomenon in nature as for the evaporation of liquids from water drops \cite{Cazabat2010} to lakes or oceans \cite{Bodenschatz2000} and in the industry, in particular for coating processes.
Indeed, the transport of solutes is often driven by evaporation and dictates the self-organization of particles at micro and macroscopic lengthscales \cite{Routh2013}.
Therefore, a correct modeling of the evaporation dynamics is crucial for understanding evaporation kinetics and colloidal deposition.

When the evaporation is limited by the diffusion in the vapor phase \cite{Langmuir1918}, the derivation of the evaporative flux shows that the total evaporation rate is linear with the radius of the liquid surface \cite{Deegan1997}.
In 2006, Shahidzadeh-Bonn \textit{et al.} reported experimental observations of evaporating drops of water and hexane \cite{Shahidzadeh-Bonn2006}.
They observed that while a drop of hexane evaporates as predicted by a diffusive model, a drop of water has an anomalous evaporation rate above a certain radius.
They attributed this different behavior to the natural convection that takes place, as water vapor is less dense than air.
The dimensionless number representing the balance between the buoyant forces in favor of convection and the viscous forces in favor of diffusion, is the Grashof number defined as
\begin{equation}\label{eq:grashof}
    {\rm Gr} = \left| \frac{\rho_s - \rho_\infty}{\rho_\infty} \right| \frac{g R^3}{\nu^2},
\end{equation}
where $\rho_s$ and $\rho_\infty$ are respectively the vapor density at saturation and at infinity, $g$ the gravity constant, $R$ the radius of the drop and $\nu$ the kinematic viscosity of air.
For small Grashof numbers, evaporation is limited by diffusion while for large Grashof numbers, buoyancy creates a convective flow.

Additional experimental questioning \cite{Dunn2009} and evidences \cite{Weon2011,Kelly-Zion2011,Kelly-Zion2013,Carle2013a,Somasundaram2015,Boulogne2017a} of the importance of convective effects have been reported more recently in the literature.
Direct visualizations of evaporating drops have been achieved by X-ray imaging method \cite{Weon2011}, IR absorption \cite{Kelly-Zion2013}, Schlieren technique \cite{Kelly-Zion2013a} or by interferometric measurements \cite{Dehaeck2014} and these studies concluded that evaporation can be enhanced by convection.
The transition between diffusive and convective evaporation regimes has been reported and the convective evaporation rate can be captured as ${\rm Gr}^\beta$ with $\beta\approx 0.20$ \cite{Kelly-Zion2011,Kelly-Zion2013,Carle2013a}.

As we expect from the definition of the Grashof number (Eq. \eqref{eq:grashof}), the threshold for convective evaporation is a function of the radial lengthscale of the evaporating surface.
Recently, Carrier \textit{et al.} investigated the mutual influence of closely deposited drops on the evaporation rate \cite{Carrier2016} as it can be encountered for a sprayed liquid.
If the drops are separated by a distance comparable to their radius, they found that a cooperative effect induces convective evaporation.
In addition, they studied the evaporation of circular evaporating surfaces of different radii and they concluded that for large Grashof numbers, convection dominates with an evaporation rate that scales as $R^2$.

    From the point of view of the governing equations, the evaporation of a liquid is similar to the dissolution of a liquid into another.
    Recent attention has been devoted to the sessile drop dissolution \cite{Zhang2015a,Dietrich2016,Laghezza2016} and bubble growth in supersaturated solutions \cite{Enriquez2014} with a combination of experimental, numerical and theoretical investigations.
    These studies show that natural convection is also observed above dissolving drops.
    The main difference between evaporation and dissolution is the magnitude of the Schmidt number $\mathrm{Sc} = \nu/{\cal D}$ defined as the ratio of the kinematic viscosity $\nu$ of the surrounding phase and the diffusivity ${\cal D}$ of the molecules.
    For evaporation, the Schmidt number is close to unity as diffusivity and kinematic viscosity are similar for gases.
    In contrast, the Schmidt number is large for dissolution \cite{Dietrich2016}.
    This difference of Schmidt number is important for the structure of the boundary layer of the convective flow \cite{Bejan1993}.

In this paper, we propose to derive scaling laws for a flat circular evaporating surface in a regime dominated by convection.
We base our analysis on an analogy between convective evaporation of flat circular surfaces and the thermal convection that takes place above a heated disk.
We show that our prediction is in agreement with our measurements and with some empirical predictions available in the literature.

\section{Model}\label{sec:model}

As stated in the introduction, the evaporation of the liquid changes the composition of the surrounding atmosphere and thus, its local density.
Therefore, we consider the natural convection that can take place above a circular disk of liquid.
In this Section, we establish a scaling law between the evaporation flux and the radius of the flat drop.
By analogy between heat and mass transfer, this derivation closely follows that of the heat transfer above a horizontal heated surface, which is a classical problem \cite{Stewartson1958,Rotem1969,Zakerullah1979,Merkin1983}.
Dehaeck \textit{et al.} have also performed a detailed study of the role of natural convection on an evaporating pendant droplet, see the Supporting Information of \cite{Dehaeck2014}. 
In particular, they discussed the role of the varying slope of the droplet, and of thermal Marangoni flows induced in the droplet by the latent heat released during evaporation.
However, these papers either considered a plate geometry instead of a disk \cite{Stewartson1958,Rotem1969}, or started from dimensionless or simplified equations \cite{Zakerullah1979,Merkin1983,Dehaeck2014}, which is why we prefer to present the derivation in details.

\begin{figure}
    \includegraphics[width=\linewidth]{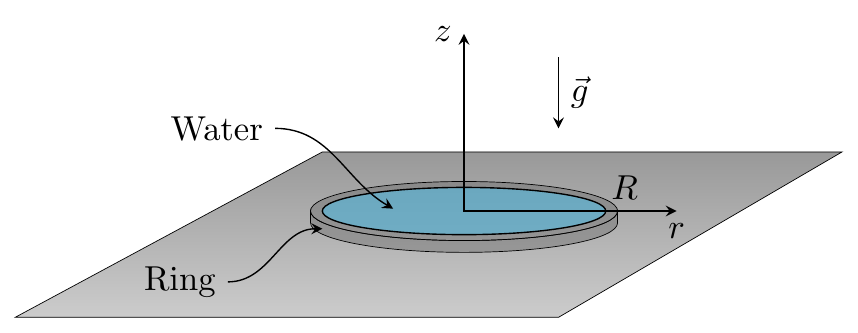}
    \caption{Sketch presenting the notations for the convective evaporation above a circular disk of volatile liquid.
    }
    \label{fig:Sketch}
\end{figure}

\subsection{Diffusion-convection equations}\label{sec:diff-conv_equations}

We establish the equations to describe the flow in the gas phase and our analysis requires the following assumptions.
(i) The liquid-vapor interface is horizontal.
(ii) The flow of vapor is axisymmetric and in a steady state.
(iii) Thermal effects are negligible.
(iv) We use the Boussinesq approximation: the air density is assumed constant, except in the terms where it acts as a driving force for the flow; in particular, the flow is considered incompressible.


Under these conditions, the velocity field of the vapor above the drop writes $\mathbf{u} = u\,\mathbf{e}_r + w\,\mathbf{e}_z$ and depends only on the cylindrical coordinates $r$ and $z$, with origin at the center of the drop interface (Fig.~\ref{fig:Sketch}).
The continuity equation is
\begin{equation}\label{eq:continuity}
    \frac{1}{r} \frac{\partial (r u) }{\partial r}  + \frac{\partial w}{\partial z} = 0.
\end{equation}
With $p$ the pressure field, $c$ the mass concentration field, $\rho(c)$ the gas density field and $\mu$ its dynamic viscosity, the radial and vertical components of the Navier-Stokes equation are respectively
\begin{subequations}\label{eq:NS1}
    \begin{align}
        \rho_\infty \left( u \frac{\partial u }{\partial r } + w \frac{\partial u }{\partial z } \right) &= -\frac{\partial p }{\partial r } +  \nonumber\\ & \mu \left( \frac{\partial^2 u }{\partial r^2} + \frac{1}{r} \frac{\partial u}{\partial r} - \frac{u}{r^2} + \frac{\partial^2 u}{\partial z^2} \right) ,\\
        \rho_\infty \left( u\frac{\partial w}{\partial r} + w\frac{\partial w}{\partial z} \right) &= -\rho(c) g - \frac{\partial p}{\partial z} + \nonumber\\ &\mu \left( \frac{\partial^2 w}{\partial r^2} + \frac{1}{r} \frac{\partial w}{\partial r} + \frac{\partial^2 w}{\partial z^2} \right). \label{Eq:Navier-Stokes_vertical_1}
    \end{align}
\end{subequations}
In these equations, according to the Boussinesq approximation, the gas density is taken as constant and equal to its value at infinity $\rho_\infty$, except in the driving force term of natural convection, $\rho(c) g$ in the equation (\ref{Eq:Navier-Stokes_vertical_1}) of vertical motion.
From the ideal gas law, the gas density varies linearly with the vapor concentration:
\begin{equation} \label{eq:rho(c)}
    \rho(c) = \rho_0 - \Delta\rho \frac{c}{c_s} ,
\end{equation}
where $\rho_0$ is the density of dry air, $\Delta\rho = \rho_0 - \rho_s$ the density difference between pure air and air saturated with vapor, and $c_s$ the saturation concentration of vapor in air.
Finally, the diffusion-convection equation for the concentration field is
\begin{equation}\label{eq:diff_conv}
    u\frac{\partial c}{\partial r} + w\frac{\partial c}{\partial z} = {\cal D} \left[ \frac{1}{r} \frac{\partial}{\partial r} \left( r \frac{\partial c}{\partial r} \right) + \frac{\partial^2 c}{\partial z^2} \right],
\end{equation}
where ${\cal D}$ is the diffusion coefficient of vapor in air.

Let $p = p_{\mathrm{stat}} + \varpi$ with $p_{\mathrm{stat}}$ the hydrostatic pressure, such that $ -\rho_\infty g - \partial p_{\mathrm{stat}}/\partial z = 0$.
The Navier-Stokes equations  \eqref{eq:NS1} become:
\begin{subequations}\label{eq:NS2}
    \begin{align}
        \rho_\infty \left( u\frac{\partial u}{\partial r} + w\frac{\partial u}{\partial z} \right) &= -\frac{\partial\varpi}{\partial r} + \nonumber\\ &\mu \left( \frac{\partial^2 u}{\partial r^2} + \frac{1}{r} \frac{\partial u}{\partial r} - \frac{u}{r^2} + \frac{\partial^2 u}{\partial z^2} \right) ,\\
        \rho_\infty \left( u\frac{\partial w}{\partial r} + w\frac{\partial w}{\partial z} \right) &= -\left[\rho(c) - \rho_\infty \right] g - \frac{\partial\varpi}{\partial z} + \nonumber\\ &\mu \left( \frac{\partial^2 w}{\partial r^2} + \frac{1}{r} \frac{\partial w}{\partial r} + \frac{\partial^2 w}{\partial z^2} \right) .
    \end{align}
\end{subequations}

We now nondimensionalize equations \eqref{eq:continuity}, \eqref{eq:diff_conv} and \eqref{eq:NS2} by the characteristic length scale $R$, the characteristic velocity $\nu/R$ and the characteristic pressure $\rho_\infty \nu^2/R^2$. We thus introduce the following dimensionless quantities, denoted by a tilde: $\tilde{r} = r/R$, $\tilde{z} = z/R$, $\tilde{u} = Ru/\nu$, $\tilde{w} = Rw/\nu$ and $\tilde{p} = R^2 p/\rho_\infty \nu^2$.
We also define a dimensionless concentration field $\tilde c = (c - c_\infty)/(c_s - c_\infty)$, where $c_\infty$ is the vapor concentration far from the drop.
We then obtain:
\begin{subequations}\label{eq:set_equations_to_solve}
    \begin{align}
        \frac{1}{\tilde r} \frac{\partial (\tilde r \tilde u) }{\partial \tilde r}  + \frac{\partial \tilde w}{\partial \tilde z} &= 0 , \label{Eq:continuite} \\
        \tilde u\frac{\partial \tilde u}{\partial \tilde r} + \tilde w\frac{\partial \tilde u}{\partial \tilde z} &= -\frac{\partial \tilde \varpi}{\partial \tilde r} + \frac{\partial^2 \tilde u}{\partial \tilde r^2} + \frac{1}{\tilde r} \frac{\partial \tilde u}{\partial \tilde r} - \frac{\tilde u}{\tilde r^2} + \frac{\partial^2 \tilde u}{\partial \tilde z^2} ,\label{Eq:Navier-Stokes_horizontal}\\
        \tilde u\frac{\partial \tilde w}{\partial \tilde r} + \tilde w\frac{\partial \tilde w}{\partial \tilde z} &= -\frac{\partial\tilde \varpi}{\partial \tilde z} + \mathrm{Gr}\,\tilde c + \frac{\partial^2 \tilde w}{\partial \tilde r^2} + \frac{1}{\tilde r} \frac{\partial \tilde w}{\partial \tilde r} + \frac{\partial^2 \tilde w}{\partial \tilde z^2} \label{Eq:Navier-Stokes_vertical},\\
        \tilde u\frac{\partial \tilde c}{\partial \tilde r} + \tilde w\frac{\partial \tilde c}{\partial \tilde z} &= \frac{1}{\mathrm{Sc}} \left[ \frac{1}{\tilde r} \frac{\partial}{\partial \tilde r} \left( \tilde r \frac{\partial \tilde c}{\partial \tilde r} \right) + \frac{\partial^2 \tilde c}{\partial \tilde z^2} \right] ,\label{Eq:convection_diffusion}
    \end{align}
\end{subequations}
with two dimensionless numbers, the Grashof number defined by equation (\ref{eq:grashof}), and the Schmidt number $\mathrm{Sc} = \nu/{\cal D}$, which is of order one because all diffusivities have the same order of magnitude in gases.

Since the liquid is out of equilibrium with the atmosphere, a mass transfer occurs at the liquid-vapor interface.
From Fick's law, the local flux normal at the liquid-vapor interface is given by
\begin{equation}\label{eq:def_local_flux}
    j(r) = -{\cal D} \left.\frac{\partial c(r, z)}{\partial z}\right|_{z = 0},
\end{equation}
where the concentration gradient is taken at the liquid-vapor interface, \textit{i.e.} $z=0$.
The total evaporating flux is  given by
\begin{equation}\label{eq:def_total_flux}
    Q = \int_S j(r)\,{\rm d} S,
\end{equation}
where $S$ is the surface of the liquid-vapor interface.

To determine these local and total fluxes, we must establish the vapor concentration gradient at the interface.
Therefore, we analyze in the next paragraphs the set of equations \eqref{eq:set_equations_to_solve} to derive this concentration gradient.

\subsection{Diffusion-limited evaporation}\label{sec:diffusion}

For small Grashof numbers, the unique driving term of the gas flow, namely $\mathrm{Gr}\,\tilde c$ in  equation (\ref{Eq:Navier-Stokes_vertical}), vanishes.
From the definition of the Grashof number given by equation (\ref{eq:grashof}), and from (\ref{eq:rho(c)}), the condition ${\rm Gr} <1$ is equivalent to a drop radius smaller than $R^\star$ defined as
\begin{equation} \label{eq:Rstar}
    R^\star = \left( \frac{c_s - c_\infty}{c_\infty} \frac{\nu^2}{g} \right)^{1/3} .
\end{equation}

\begin{figure}
	\centering
    \includegraphics[width=1\linewidth]{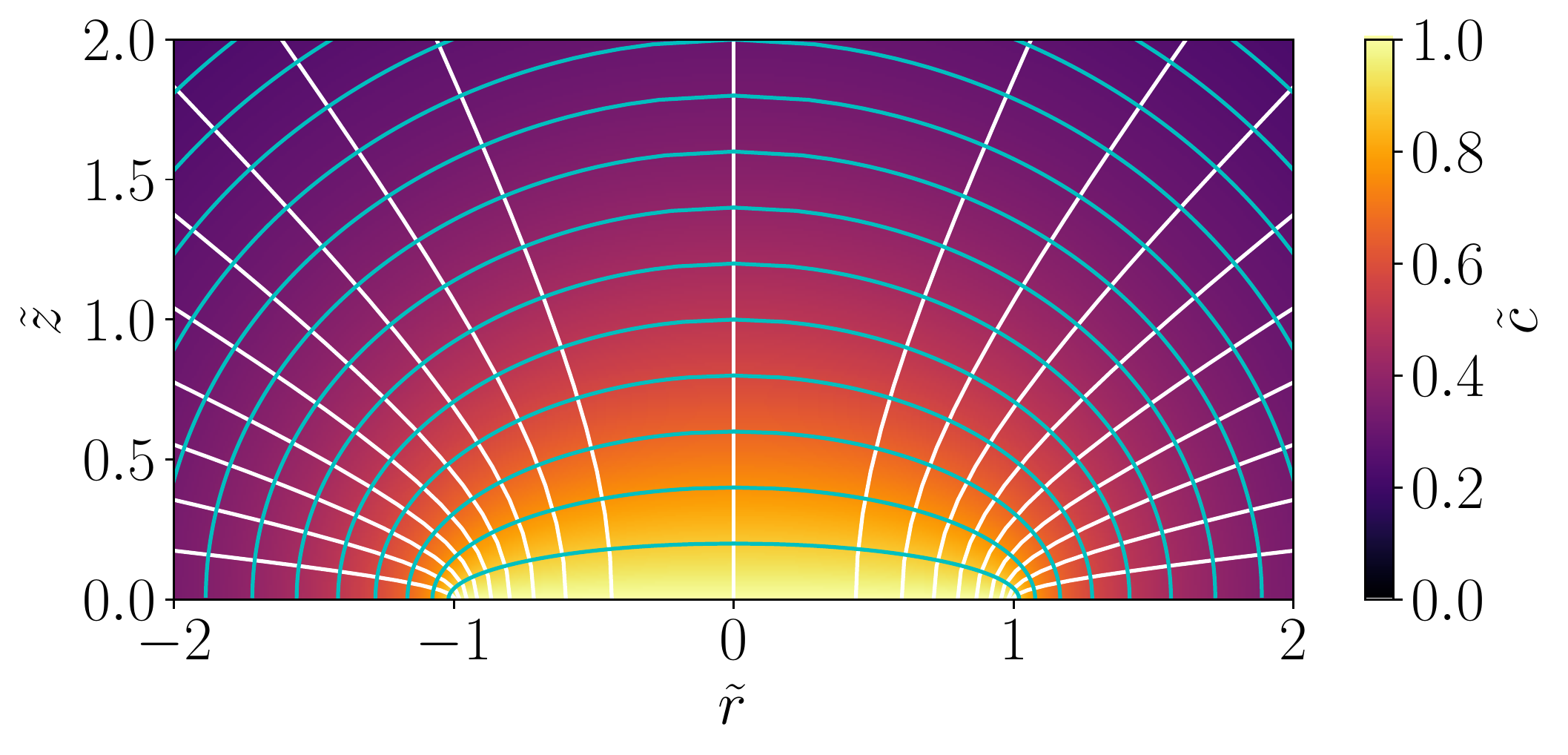}
    \caption{Dimensionless vapor concentration map above an evaporating disk in the dimensionless space $(\tilde r, \tilde z)$ obtained from equation \eqref{eq:vapor_disk}. 
    White and cyan lines are the oblate spheroidal coordinates $(\kappa, \sigma)$, for respectively constant $\kappa$ and $\sigma$ values.}
    \label{fig:marder}
\end{figure}

Hence, in this condition, there is no gas flow \textit{i.e.} $\tilde u=\tilde w=0$, and the set of equations (\ref{eq:set_equations_to_solve}) reduces to the Laplace equation for the concentration field.
From equation \eqref{Eq:convection_diffusion}, the Laplace equation writes

\begin{equation} \frac{1}{\tilde r} \frac{\partial}{\partial \tilde r} \left( \tilde r \frac{\partial \tilde c}{\partial \tilde r} \right) + \frac{\partial^2 \tilde c}{\partial \tilde z^2} = 0.
\end{equation}
The boundary conditions are a saturated vapor concentration at the surface of the disk $c(r<R, z=0) = c_s$ and a vapor concentration $c_\infty$ far from the disk.
Introducing the oblate spheroidal coordinates ($\kappa, \sigma$) defined as $\tilde{r}^2 = (1-\kappa^2)(1+\sigma^2)$ and $\tilde{z}= \kappa\sigma$, the vapor concentration can be written in the simple form \cite{Lebedev1965,Marder1981}
\begin{equation}\label{eq:vapor_disk}
    \tilde c(\kappa, \sigma)   =  1- \frac{2}{\pi} \arctan(\sigma) .
\end{equation}
This solution is represented in Fig.~\ref{fig:marder}.

Thus, for a diffusion-limited evaporation, the local flux calculated from equation \eqref{eq:def_local_flux}, is
\begin{equation}\label{eq:local_flux_diffusive}
    j_{\rm diff}(r) = \frac{2}{\pi} \frac{{\cal D} (c_s - c_\infty) }{\sqrt{R^2 - r^2}},
\end{equation}
which is the well-known flux for a drop of a small contact angle.
This flux presents a divergence at the edge that can be interpreted as a tip effect by analogy with electrostatic problems.
Substituting equation \eqref{eq:local_flux_diffusive} in equation \eqref{eq:def_total_flux}, we derive the total flux
\begin{equation}\label{eq:total_flux_diffusive}
    Q_{\rm diff} = 4 {\cal D} (c_s - c_\infty) R.
\end{equation}

Equations (\ref{eq:local_flux_diffusive}) and (\ref{eq:total_flux_diffusive}) have been largely commented in the literature, particularly in the frames of evaporating droplet lifetime \cite{Stauber2014,Stauber2015a} and the so-called coffee stain effect \cite{Deegan1997,Deegan2000a,Popov2005,Marin2011,Boulogne2017a}.
Therefore, we do not develop further the diffusion-limited case and we analyze in the next paragraph the convective evaporation.

\subsection{Convective evaporation}\label{sec:convection}

\begin{figure}
    \includegraphics[width=\linewidth]{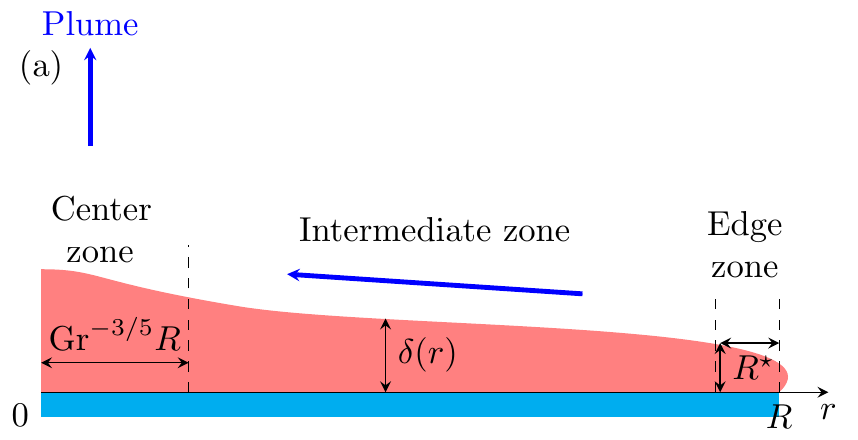}\\
    \includegraphics[width=\linewidth]{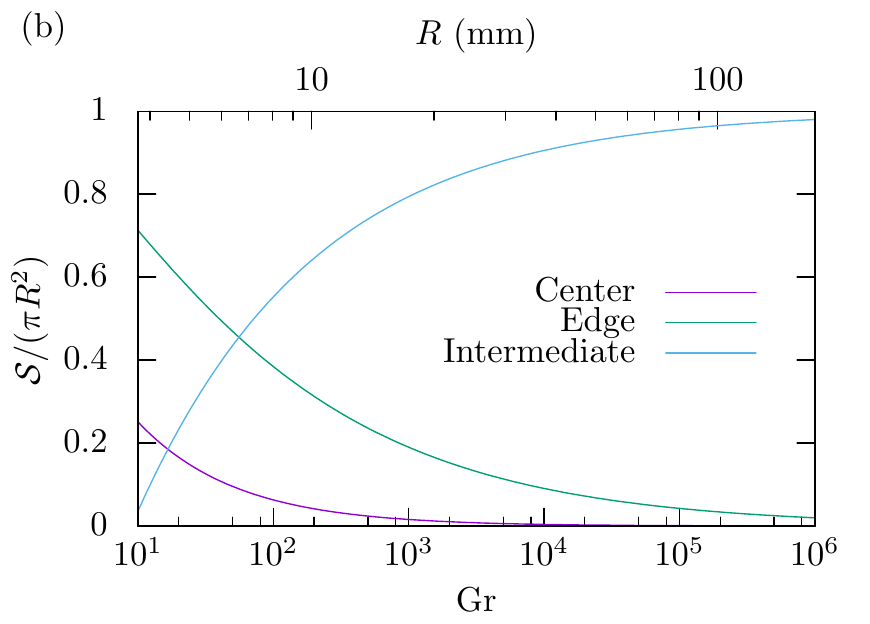}
    \caption{
    (a) For a large Grashof number, sketch representing the three characteristic zones of the boundary layer depicted in red above an evaporating disk of liquid.
    The thickness of the boundary layer is given by Eq.~ \eqref{eq:boundary_layer_thickness}.
    Blue arrows indicate the direction of the gas flow for a light vapor, $\Delta \rho >0$.
    (b) Relative surface area of the center, edge and intermediate zone as a function of the Grashof number.
    Each area is calculated from the scaling presented in (a).
    The corresponding radius $R$ is indicated for water.}
    \label{fig:theo_flux}
\end{figure}

\subsubsection{Spatial structure of the boundary layer}

For large Grashof numbers, a flow is established in the gas phase.
The direction and the spatial structure of this flow depends on the sign of the density difference $\Delta\rho$, as well as the spatial structure of the flow, especially at the edge and at the center of the liquid disk.
As the experiments conducted in Section~\ref{sec:experiments} are for a positive density difference, we henceforth assume that $\Delta \rho>0$.
Under this assumption, the flow is mostly horizontal inwards.
Due to this convective flow, unsaturated air is brought toward the edge of the evaporating film.
At some distance from the edge of the film, a slender boundary layer is established and progressively thickens.
At the center of the disk, the flow field converges, and thus moves upwards to form a rising plume of wet air. 

Consequently, we distinguish three zones as depicted in Fig. \ref{fig:theo_flux}(a):
the \emph{edge} and the \emph{center} zones, that do not satisfy the conditions for a slender boundary layer, in contrast to the \emph{intermediate} zone, which is slender in the sense that vertical variations are much sharper than horizontal ones.
We first focus on this intermediate zone where the approximation of a thin horizontal boundary layer can be applied to determine its thickness variation.
Then, from this first analysis, we precise the horizontal extension of the intermediate zone and, by consequence, we also define the sizes of the edge and center zones.

\subsubsection{Intermediate zone}

We assume that the flow is almost horizontal and that the vertical variations occur over lengthscales mush smaller than the horizontal ones.
To retain the necessary inertial, viscous and pressure terms describing such a flow, the following rescaling is introduced \cite{Stewartson1958}:
$\hat{z} = \mathrm{Gr}^{1/5}\tilde z$, $\hat{u} = \mathrm{Gr}^{-2/5}\tilde u$, $\hat{w} = \mathrm{Gr}^{-1/5}\tilde w$ and $\hat{\varpi} = \mathrm{Gr}^{-4/5} \tilde \varpi$.
Substituting these new variables in equations (\ref{eq:set_equations_to_solve}), we obtain:
\begin{subequations}
    \begin{align}
        \frac{1}{\tilde r} \frac{\partial (\tilde r\hat{u}) }{\partial \tilde r} + \frac{\partial\hat{w}}{\partial\hat{z}} =& 0 ,\\
        \hat{u} \frac{\partial\hat{u}}{\partial \tilde r} + \hat{w} \frac{\partial\hat{u}}{\partial\hat{z}} =& -\frac{\partial\hat{\varpi}}{\partial \tilde r} + \nonumber\\ & \mathrm{Gr}^{-2/5} \left( \frac{\partial^2 \hat{u}}{\partial \tilde r^2} + \frac{1}{\tilde r} \frac{\partial\hat{u}}{\partial \tilde r} - \frac{\hat{u}}{\tilde r^2} \right) \nonumber \\ &+ \frac{\partial^2 \hat{u}}{\partial\hat{z}^2} ,\\
        \mathrm{Gr}^{-2/5} \left( \hat{u} \frac{\partial\hat{w}}{\partial \tilde r} + \hat{w} \frac{\partial\hat{w}}{\partial\hat{z}} \right) =& -\frac{\partial\hat{\varpi}}{\partial\hat{z}} + \tilde c + \nonumber\\ & \mathrm{Gr}^{-4/5} \left( \frac{\partial^2 \hat{w}}{\partial \tilde r^2} + \frac{1}{\tilde r} \frac{\partial\hat{w}}{\partial \tilde r} \right) \nonumber\\
        &+ \mathrm{Gr}^{-2/5} \frac{\partial^2 \hat{w}}{\partial\hat{z}^2} ,\\
        \hat{u} \frac{\partial \tilde c}{\partial \tilde r} + \hat{w} \frac{\partial \tilde c}{\partial\hat{z}} = \frac{1}{\mathrm{Sc}} &\left[ \mathrm{Gr}^{-2/5} \frac{1}{\tilde r} \frac{\partial}{\partial \tilde r} \left( \tilde r \frac{\partial \tilde c}{\partial \tilde r} \right) + \frac{\partial^2 \tilde c}{\partial\hat{z}^2} \right].
    \end{align}
\end{subequations}
Hence, within corrections of relative order $\mathrm{Gr}^{-2/5}$,
\begin{subequations} \label{Eq:equations_couche_limite}
    \begin{align}
        \frac{1}{\tilde r} \frac{\partial (\tilde r\hat{u})}{\partial \tilde r}  + \frac{\partial\hat{w}}{\partial\hat{z}} &= 0 , \label{Eq:continuite_couche_limite} \\
        \hat{u} \frac{\partial\hat{u}}{\partial \tilde r} + \hat{w} \frac{\partial\hat{u}}{\partial\hat{z}} &= -\frac{\partial\hat{\varpi}}{\partial \tilde r} + \frac{\partial^2 \hat{u}}{\partial\hat{z}^2} , \label{Eq:Navier-Stokes_horizontal_couche_limite} \\
        0 &= -\frac{\partial\hat{\varpi}}{\partial\hat{z}} + \tilde c , \label{Eq:Navier-Stokes_vertical_couche_limite} \\
        \hat{u} \frac{\partial \tilde c}{\partial \tilde r} + \hat{w} \frac{\partial \tilde c}{\partial\hat{z}} &= \frac{1}{\mathrm{Sc}} \frac{\partial^2 \tilde c}{\partial\hat{z}^2} , \label{Eq:convection_diffusion_couche_limite}
    \end{align}
\end{subequations}
which are the analogue equations that describe natural convection above a heated disk \cite{Merkin1983}.

Although these equations can only be solved numerically, the full solution is not necessary to obtain the scaling of the evaporative flux.
Starting from the edge of the drop ($\tilde r = 1$), a boundary-layer solution emerges.
To see this, we set $\tilde r = 1 - \tilde x$ in equations (\ref{Eq:equations_couche_limite}), and consider their behavior at small $\tilde x$.
We then obtain equations similar to those describing natural convection above a horizontal heated plate with a straight edge.
As shown by Stewartson \cite{Stewartson1958}, such equations admit a self-similar solution depending on the rescaled variable\footnote{Notice that there is a typo in the definition (9) of $\eta$ in \cite{Stewartson1958}.} $\hat{\eta} = \hat{z}/\tilde x^{2/5}$.
Merkin \cite{Merkin1983} showed that, although such a self-similarity is lost when considering equations (\ref{Eq:equations_couche_limite}), it is still possible to use $\hat{\eta}$ as a rescaling variable for $\hat{z}$. Hence, the concentration gradient is localized within a boundary layer of dimensionless thickness $\tilde x^{2/5}$. In dimensional units, from the definitions of $\hat{z}$, $\tilde{z}$ and $\tilde{x}$, the scaling of the boundary layer thickness is thus
\begin{equation}\label{eq:boundary_layer_thickness}
    \delta(r) \approx \frac{(R-r)^{2/5} R^{3/5}}{ \mathrm{Gr}^{1/5}}.
\end{equation}

\subsubsection{Spatial extension of the three zones}

Now, we estimate the spatial extension of the different zones.
From equation \eqref{eq:boundary_layer_thickness}, the slenderness of the boundary layer scales as
\begin{equation} \label{Eq:slenderness}
    \frac{\delta}{x} = \frac{1}{\mathrm{Gr}^{1/5}} \left(\frac{R}{x}\right)^{3/5} .
\end{equation}
As a consequence, starting from the drop edge, the boundary layer becomes slender only for distances larger than $x = R/\mathrm{Gr}^{1/3} = R^\star$ from Eqs.~(\ref{eq:grashof}) and (\ref{eq:Rstar}), which defines the extent of the edge zone; in this domain, horizontal and vertical variations occur over a similar lengthscale $R^\star$.
This is similar to the breakdown of the boundary layer approximation, for the flow of a fluid at high Reynolds number, at the immediate vicinity of the leading edge of a solid, which is well known in fluid mechanics \cite{Leal2007}.
This effect is often negligible in the estimation of the friction force on a solid.
However, as we discuss later, the contribution of the edge on the total evaporation rate is significant in our situation.

Close to the center of the drop, the solution given by Eq.
\eqref{eq:local_flux_convective} also breaks down, and must match the
rising plume.
As discussed by Merkin \cite{Merkin1983}, the situation is
complex close to the center, where a strong pressure builds up in response to the
converging horizontal flow.
Indeed, the flow structure is still described by
Eqs.~\eqref{Eq:equations_couche_limite} but, approaching the center, the
boundary layer splits into two layers: an inner layer in contact with
the drop where viscous and inertial effects balance the pressure buildup
while concentration gradients are not significant, and an outer inviscid
layer where the concentration gradients are significant.
The inner layer has a dimensionless thickness $\tilde r^{2/3}$ \cite{Merkin1983}.
Similarly to equation \eqref{Eq:slenderness}, the slenderness of the inner layer scales as $\mathrm{Gr}^{-1/5} (R/r)^{1/3}$.
Thus, the inner layer cannot be considered as slender for $r \lesssim
\mathrm{Gr}^{-3/5} R$, which defines the extent of the central zone
where the boundary-layer approximation underlying equation
\eqref{Eq:equations_couche_limite} breaks down (Fig.
\ref{fig:theo_flux}(a)).

\subsubsection{Evaporative flux}
A proper estimate of the evaporative flux must consider in principle the three zones (Fig. \ref{fig:theo_flux}(a)).
In the edge zone $x < R^\star$, we showed that the boundary layer has both horizontal and vertical extensions of lengthscale $R^\star$.
Thus, we can estimate the scaling of the local flux defined by equation \eqref{eq:def_local_flux} as
\begin{equation}
	j_{\rm edge} \approx {\cal D} \frac{ c_s - c_\infty}{{R^\star}},
\end{equation}
where ${R^\star}$ is the characteristic lengthscale of the vertical vapor concentration gradient.
The domain $x < R^\star$ has a surface $2\pi RR^\star$  in the limit of large Grashof numbers, \textit{i.e.} $R^\star \ll R$.
Hence, the evaporative flux in this zone has the following order of magnitude:
\begin{equation}\label{eq:flux_edge}
    Q_{\rm edge} \approx	2\pi RR^\star  \frac{ {\cal D}(c_s - c_\infty) }{R^\star} = 2\pi {\cal D}R(c_s - c_\infty).
\end{equation}
Interestingly, this is exactly the same scaling as the evaporation flux of the whole drop in the purely diffusive regime as shown in Eq. \eqref{eq:total_flux_diffusive}.

In the intermediate zone, for $r < R - R^\star$, the flux scales as
\begin{equation}
    j_{{\rm int}}(r) \approx  {\cal D}  \frac{c_s - c_\infty}{\delta(r)},
\end{equation}
where $\delta(r)$ is the boundary layer thickness that corresponds to the lengthscale of the vapor concentration gradient.
From the scaling of $\delta(r)$ given by Eq. \eqref{eq:boundary_layer_thickness}, we have
\begin{equation}\label{eq:local_flux_convective}
    j_{{\rm int}}(r)  \approx \frac{{\cal D} (c_s - c_\infty)  {\rm Gr}^{1/5} }{R^{3/5} (R-r)^{2/5}} ,
\end{equation}
and using equation (\ref{eq:def_total_flux}) with ${\rm d} S = 2\pi r {\rm d} r$, we obtain
\begin{equation}\label{eq:scaling_convective_flow_rate}
    Q_{{\rm int}} \approx  2\pi {\cal D}R(c_s - c_\infty)\mathrm{Gr}^{1/5} = 2\pi\frac{{\cal D}(c_s - c_\infty)^{6/5} }{c_\infty^{1/5} \nu^{2/5}} R^{8/5} .
\end{equation}
The main prediction of this analysis is that the evaporation flux must be proportional to $R^{8/5}$.

A corrective term must be established for the center zone.
However, the ratio of the central zone area to the total drop area is $\mathrm{Gr}^{-6/5}$.
As shown in Fig. \ref{fig:theo_flux}(b), the center zone represents a small portion of the drop.
Thus, we neglect the contribution of the center zone on the total evaporative flux.

Consequently, from Eqs. \eqref{eq:scaling_convective_flow_rate} and \eqref{eq:flux_edge}, our analysis yields the following law for the evaporation flux:
\begin{subequations}\label{eq:total_convective_flow_rate}
    \begin{align}
    Q_{\rm conv} &\approx a_1 Q_{{ \rm int}} + a_2 Q_{{\rm edge}} \\
    &\approx 2\pi {\cal D}R(c_s - c_\infty)(a_1 \mathrm{Gr}^{1/5} + a_2)\label{eq:total_convective_flow_rateB},
        \end{align}
\end{subequations}
where $a_1$ and $a_2$ are two constant numbers.
By inspection of Eq.~(\ref{eq:total_convective_flow_rateB}), we notice that in a convective regime, the total flux is a combination of a term reminiscent of a diffusive evaporation regime and a second term depending on the Grashof number, a signature of the convective flow.
In the following, we perform experiments to measure the evaporation rate for large Grashof numbers to validate our prediction given by equation \eqref{eq:total_convective_flow_rate}.

\section{Experiments}\label{sec:experiments}

\begin{figure}
    \includegraphics[width=\linewidth]{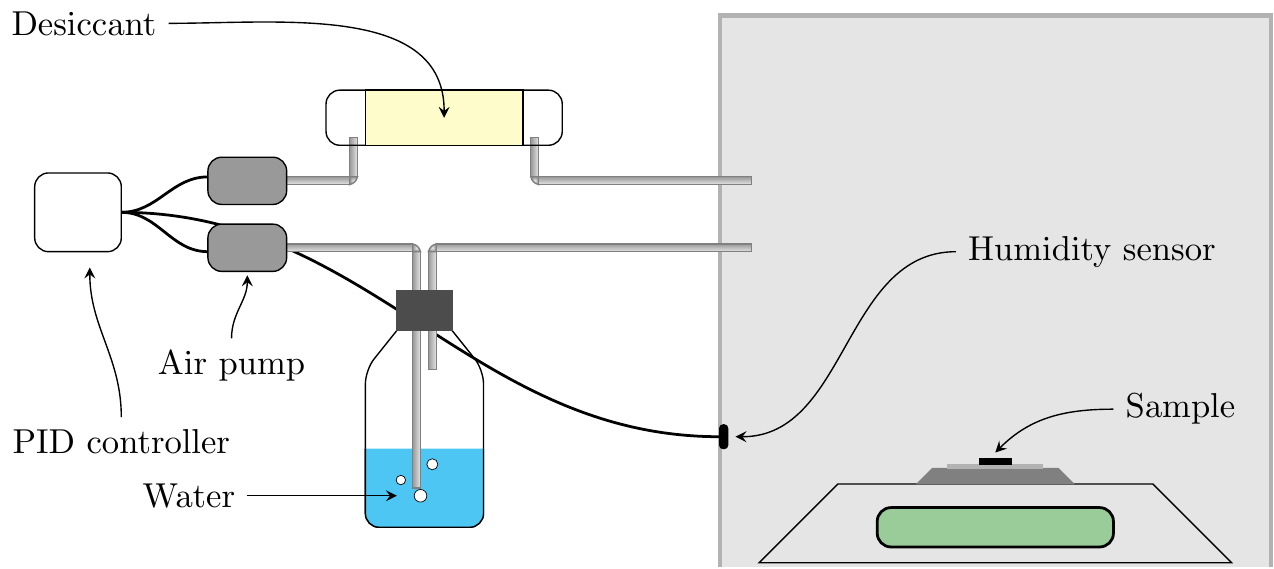}
    \caption{ Sketch of the experimental setup with a scale in a box for which the atmosphere is controlled in humidity.
    }
    \label{fig:Setup}
\end{figure}

        Our experiments of controlled evaporation are performed in a box made in polycarbonate ($50\times50\times50$ cm$^3$).
        A precision scale (Ohaus Pioneer 210~g) with a precision of $0.1$ mg is placed at the center of the box and is interfaced with a Python code using the \textit{pyserial} library to record the time evolution of the weight.
The humidity is regulated with a PID controller based on an Arduino Uno and a humidity sensor (Honeywell HIH-4021-003) positioned far from the evaporating surface (Fig. \ref{fig:Setup}).
        Dry air is produced by circulating ambient air with an air pump (Tetra APS 300) in a container filled with desiccant made of anhydrous calcium sulfate (Drierite).
        Moist air is obtained by bubbling air in water.
        The relative humidity is set to $R_H=50$\% in all of our experiments.

\begin{figure}
    \includegraphics[width=0.98\linewidth]{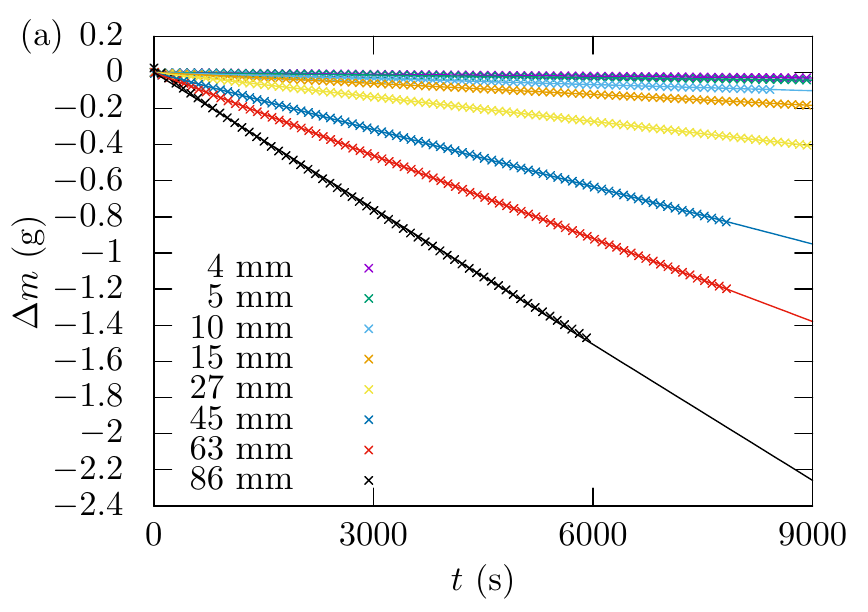}\\
    \includegraphics[width=0.98\linewidth]{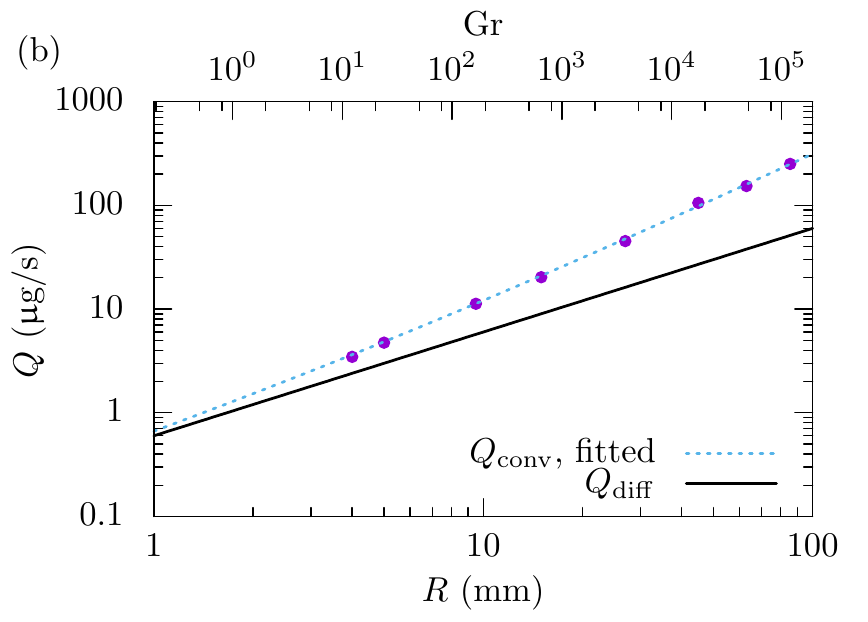}\\
    \includegraphics[width=0.98\linewidth]{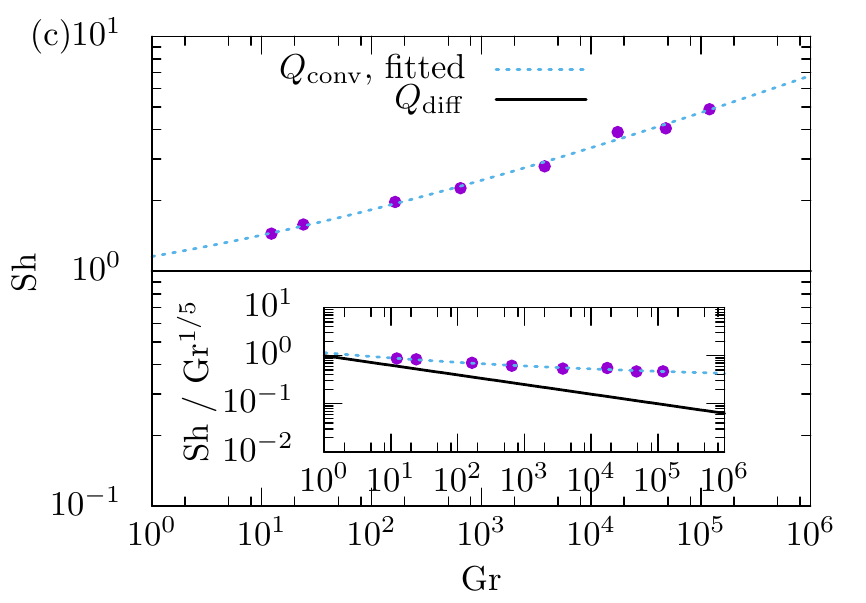}\\
    \caption{(a) Time evolution of the weight of water for different radii of circular troughs.
    The solid lines corresponds to linear fit of the experimental data points.
    (b) Measured evaporation rate as a function of the surface radius and the Grashof number.
    The dotted blue line is a fit with equation \eqref{eq:total_convective_flow_rate} with a Grashof number to the power $0.18$, and with the coefficients $a_1 =0.31$ and $a_2=0.48$.
    The solid line is the diffusive prediction $Q_{\rm diff}$.
    (c) Sherwood number ${\rm Sh}$ defined by equation \eqref{eq:sherwood} as a function of the Grashof number. Lines are equivalent to the plot (b). 
    The inset shows the compensated plot ${\rm Sh} / {\rm Gr}^{1/5}$ vs ${\rm Gr}$.}
    \label{fig:Experiments}
\end{figure}

Circular troughs of different radii are filled with pure water right to the brim and a particular attention is paid to get a flat surface.
The diffusion coefficient of water vapor in air is ${\cal D} \approx 2\times 10^{-5}$ m$^2$/s, and the kinematic viscosity of the gas is $\nu \approx 1.5 \times 10^{-5}$ m$^2$/s in our experimental conditions \cite{Boulogne2017a}.
The time variation of the weight of these containers is reported in Fig. \ref{fig:Experiments}(a) over a time duration between $6000$ to $9000$ s.
Data points are fitted with a linear function to get the mass evaporation rate $Q$ such that $\Delta m(t) = - Q t$.
In Fig. \ref{fig:Experiments}(b), we report the evaporation rates $Q$ as a function of the radius of the liquid patch and the equivalent Grashof number as defined by equation \eqref{eq:grashof}.

To estimate the Grashof number, we evaluate the density gradient in the atmosphere due to the difference of vapor concentration.
The air density $\rho$ at a pressure $P_0$, a temperature $T$ and a relative humidity $R_H$ is given by \cite{Tsilingiris2008}
\begin{equation}
    \rho = \frac{1}{z_m} \frac{P_0}{{\cal R} T} M_d \left[1 - f R_H \left( 1 - \frac{M_w}{M_d}  \right)  \frac{P_s}{P_0} \right],
\end{equation}
where $z_m$ and $f$ are the compressibility and enhancement factors, ${\cal R}$ the ideal gas coefficient, $M_d$ and $M_w$ the molar density of dry and saturated air, and $P_s$ the saturated pressure.
At room temperature, we have $f = 1$, $z_m = 1$ \cite{Tsilingiris2008} and a saturated pressure $P_s = 2.3$~kPa \cite{Tennent1971}.
Therefore, the density variation at a relative humidity $R_H=0.5$ is $|\rho_s - \rho_\infty| / \rho_\infty \approx 5 \times 10^{-3}$.
In the next section, we discuss the evaporation model that we proposed in Section \ref{sec:model} in comparison with our experimental results.

\section{Discussion}
In Fig. \ref{fig:Experiments}(b), we show a fit from equation  \eqref{eq:total_convective_flow_rateB} that correctly captures our data.
From our experimental results, we determine the values of $a_1$ and $a_2$ by fitting equation \eqref{eq:total_convective_flow_rateB} with a variable Grashof exponent.
    The resulting exponent is 0.18 and coefficient values are $a_1 =0.31$ and $a_2=0.48$ (Fig.~\ref{fig:Experiments}b).
    The two dimensionless coefficients $a_1$ and $a_2$ are close to unity and the power of the Grashof number is close to $1/5$, in agreement with the theoretical prediction.
    The fact that $a_1 \simeq a_2$ and ${\rm Gr}^{1/5}$ is between 1 and 10, shows that the contribution of the edge ($a_2\,Q_{{\rm edge}}$), although smaller, remains significant compared to the intermediate zone ($a_1\,Q_{{\rm int}}$) in our experimental conditions.

To non-dimensionalize the convective flux, we introduce the Sherwood number ${\rm Sh}$ defined as the ratio of the convective and diffusive fluxes,
\begin{equation}\label{eq:sherwood}
	{\rm Sh} = \frac{Q_{{\rm conv}}}{Q_{ {\rm diff }}} \approx \frac{\pi}{2} \left( a_1 {\rm Gr}^{1/5} + a_2 \right),
\end{equation}
from equations (\ref{eq:total_flux_diffusive}) for $Q_{ {\rm diff }}$ and (\ref{eq:total_convective_flow_rateB}) for $Q_{ {\rm conv }}$.
In Fig.~\ref{fig:Experiments}c, we plot the Sherwood number as a function of the Grashof number.
Thus, we can estimate for instance that convection becomes significant for ${\rm Sh=1.5}$, \textit{i.e.} the convective flux is 50\% higher than the predicted diffusive flux, which corresponds to ${\rm Gr} \approx 20$.

The evaporation regime in the experiments is convective.
To follow the first data analysis performed by Kelly-Zion \textit{et al.} \cite{Kelly-Zion2011}, we fit our data with a power law against the disk radius.
We obtain that $Q\propto R^{1.39}$, in agreement with their experimental results on heptane and 3-methylpentane where  the radius exponents are $1.37$ and $1.43$, respectively.
Nevertheless, we must underline that the vapors of heptane and 3-methylpentane are denser than air, contrary to water vapor.
This difference in vapor density can change the total evaporation rate, especially at the edge of the disk.
The similar values for the exponent can be attributed to the fact that this scaling mainly probes the effect of the radius due to the intermediate zone, which is insensitive to the sign of $\Delta\rho$.

In their study, Kelly-Zion \textit{et al.} \cite{Kelly-Zion2011} also considered that the total evaporation flux is the result of two significant contributions, one based on diffusion, and the other on convection.
In our theoretical analysis in Section \ref{sec:model}, we saw that the evaporative flux is always a diffusive flux evaluated right at the interface of the evaporating drop.
However, it turns out that the analysis of the spatial structure of the flow of vapor above the drop, and its separation into two convective-diffusive zones: one slender far from the edge, and one more isotropic close to the edge, justifies that the evaporative flux separates into two contributions, one scaling as a purely diffusive flux, the other one as a convective one in a regime of large Grashof number. Kelly-Zion et al. \cite{Kelly-Zion2011} studied volatile compounds with vapor heavier than air, contrary to water vapor. Although this leads to a different flow structure, starting from the center outwards, this does not affect the scaling derived in Sec.~\ref{sec:model}.
Therefore, our model justifies \textit{a posteriori} the hypothesis made by Kelly-Zion et al. \cite{Kelly-Zion2011} and later by Carle \textit{et al.} \cite{Carle2013a}, who have added a diffusive and a convective flux to model their data.

Carrier \textit{et al.} investigated the evaporation of water with beakers filled to the rim and they reported two regimes \cite{Carrier2016}.
For radii smaller than $30$ mm, the evaporation rate is proportional to the radius $R$.
For larger radii, they suggest that the evaporation rate scales as $R^2$, for radii up to $300$ mm and they concluded that this second regime can be attributed to convection.
This power exponent is significantly larger than the value found by Kelly-Zion \textit{et al.} \cite{Kelly-Zion2011} and ours.
Carrier \textit{et al.} attributed the difference to measurements by Kelly-Zion \textit{et al.} to a crossover between the diffusive and a regime characterized by the development of convection cells.
In the spirit of the classical Rayleigh-B\'enard instability, these authors proposed a direct transition from a quiescent state for the gas with diffusive exchanges solely, to an unstable convection state where the gas would be set in motion along convection cells.
However, such a transition appears in configurations where the gas would be confined between horizontal surfaces, which is far from our experimental configuration (Fig.~\ref{fig:Setup}).
Unfortunately, the precise configuration and the confinement are not reported for the experiments performed by Carrier \textit{et al.}, so we cannot conclude as to whether the difference in exponent can be ascribed to a difference in confinement.
In our case, the steady single-plume situation becomes unstable above a certain Grashof number, with the appearance of multiple plumes and even of a large-scale circulation \cite{Grossmann2000}.
However, to the best of our knowledge, the onset of such an instability is unknown.
Nevertheless, our measurements suggest that, even if such an instability appears in our range of Grashof numbers, it does not lead to a significant deviation from the scaling (\ref{eq:total_convective_flow_rate}).

\section{Conclusion}

In this paper, we used an analogy between thermal convection that occurs above a uniformly heated disk and the convective evaporation of a circular patch of liquid.
The dimensionless form of the flow equations in the gas phase reveals a dimensionless parameter called the Grashof number ${\rm Gr}$, which relates the opposite effects of buoyant and viscous forces.
This number increases with the radius $R$ of the disk as $R^3$.
Therefore, natural convection is expected for large Grashof numbers.

In our theoretical analysis, we recalled the vapor concentration field surrounding an evaporating disk and the related evaporative flux for small Grashof numbers, \textit{i.e.} a diffusion-limited evaporation.
In addition, we investigated the evaporation dynamics under natural rising convection.
The evaporating surface must be split in three domains.
Near the edge of the disk, over a distance $R^\star$, the boundary layer is not slender and the evaporative flux is diffusive with a concentration gradient across a distance $R^\star$.
In an intermediate zone, the boundary layer can be considered as slender with a thickness that scales as $(R-r)^{2/5}$ and the evaporative flux scales as ${\rm Gr^{1/5}}$.
Near the center of the disk, the flow is converging and is directed upward with a more complex boundary layer structure.
The characteristic area ratio of this zone is ${\rm Gr}^{-6/5}$, such that the precise contribution of the center zone may be neglected in a first approach, although further experimental and theoretical investigation of this zone could be performed to clarify its fine structure.
Therefore, the total evaporative flux under natural convection results in a combination of a diffusive-like flux at the edge and a flux across the boundary layer in the intermediate zone, and is correctly described by equation \eqref{eq:total_convective_flow_rate}.
The Sherwood number, which compares the convective flux $Q_{\rm conv}$ to the diffusive flux $Q_{\rm diff}$ shows that the convective flow has a significant effect for ${\rm Gr > 20}$.

With our experimental measurements, we successfully verify our prediction under natural convection.
The scaling laws of our theoretical developments provide the different evaporating mechanisms that occur above a circular liquid disk.
In particular, we explain the origin of the convective evaporation dynamics observed experimentally \cite{Kelly-Zion2011,Carle2013a}, where a combination of diffusion and convection-like contributions has been noticed.

Further refinements would require numerical calculations to provide a full resolution of the gas flow and, therefore, a prediction of the evaporation rate without unknown prefactors.
In addition, it would be interesting to investigate also the case of drops with a non-zero contact angle to evaluate how the interface slope can modify the spatial extension of the intermediate zone.
Furthermore, it would be interesting to revisit the case of vapor heavier than air, for which the flow is expected to be outwards above the evaporating disk. 
In such a case, we expect a significant effect of the geometrical configuration at the edge. 
If the disk is placed at height, a downwards gravity current starting from the edge can evacuate the vapor collected from the disk, playing the same role as the rising plume in the case of light vapor. 
In contrast, if the disk is placed on an infinite plane or in a trough, vapor cannot be evacuated, and accumulates, probably leading to a reduction of the evaporation rate.
From a general perspective, the convective flow strongly depends on the geometrical aspects, which can lead to complex expressions of the evaporative flux with effects of boundaries.

\section{Appendix}

In Section \ref{sec:diff-conv_equations}, we assume that the flow in the gas phase is steady in order to neglect the time derivatives that would appear in equations \eqref{eq:NS1}.
This assumption can be justified as follows.
For large Grashof numbers, the characteristic velocity in the gas phase is
\begin{equation}
    {\cal U} = \frac{\nu}{R} {\rm Gr}^{2/5},
\end{equation}
as suggested by the rescaling $\hat u = {\rm Gr}^{-2/5} \tilde u$ with $\tilde u=R u / \nu$.
Therefore, the characteristic timescale associated to the natural convection is \begin{equation}
    \tau_{\rm conv} = \frac{R}{ {\cal U} } = \frac{R^2}{\nu} {\rm Gr}^{-2/5}.
\end{equation}
For a disk radius of water of $R=20$ mm, the Grashof number is ${\rm Gr} \approx 1.5 \times 10^3$ and the timescale is $\tau_{\rm conv} \approx 1.4$ s, which is much smaller than the observation timescales.

Equivalently, we can estimate the timescale for small Grashof numbers.
For a diffusion-limited evaporation, the timescale is simply \cite{Crank1975}
\begin{equation}
    \tau_{\rm diff} = \frac{R^2}{{\cal D}}.
\end{equation}
For a millimeter water drop radius, $\tau_{\rm diff} \approx 5 \times 10^{-2}$ s, which is also much smaller than the lifetime of such droplet \cite{Dunn2009}.
Consequently, this supports the steady state assumption made in Section \ref{sec:diff-conv_equations}.

\section*{Acknowledgments}
We thank F. Ingremeau and H.A. Stone for stimulating discussions.
F.B. thanks M. Decraene for her assistance.
This study has been carried out with a funding support by the ANR (ANR-11-BS04-0030-WAFPI project).

        \bibliography{biblio}

        \bibliographystyle{unsrt}

        \end{document}